\documentclass[reprint, prd, superscriptaddress]{revtex4-2}
\usepackage{amsmath, amssymb}
\usepackage{graphicx,color,float,tabularx,siunitx,multirow}
\usepackage[caption=false]{subfig}
\usepackage{placeins}
\usepackage{rviewport}
\usepackage{orcidlink}
\definecolor{colorLink}{rgb}{0,0,180} 
\usepackage{hyperref}
\hypersetup{
   colorlinks = true,
   citecolor  = colorLink,
   urlcolor   = colorLink,
   linkcolor  = colorLink,
}

\DeclareSIUnit\year{yr}

\usepackage[sort&compress]{natbib}

\usepackage[shortcuts]{extdash}

\DeclareMathOperator{\sgn}{sgn}

\begin{document}

\title{Prospects for joint cosmic ray and neutrino constraints\\ on the evolution of trans-GZK proton sources}

\author{Marco~Stein~Muzio \orcidlink{0000-0003-4615-5529}}
\email{msm6428@psu.edu}
\affiliation{Department of Physics, Pennsylvania State University, University Park, PA 16802, USA}
\affiliation{Department of Astronomy and Astrophysics, Pennsylvania State University, University Park, PA 16802, USA}
\affiliation{Institute of Gravitation and the Cosmos, Center for Multimessenger Astrophysics, Pennsylvania State University, University Park,
PA 16802, USA}

\author{Michael~Unger \orcidlink{0000-0002-7651-0272}}
\email{michael.unger@kit.edu}
    \affiliation{Institute for Astroparticle Physics, Karlsruhe Institute of Technology (KIT), Karlsruhe, Germany}
    \affiliation{Institutt for fysikk, Norwegian University of Science and Technology (NTNU), Trondheim, Norway}

\author{Stephanie~Wissel \orcidlink{0000-0003-0569-6978}}
\email{wissel@psu.edu}
\affiliation{Department of Physics, Pennsylvania State University, University Park, PA 16802, USA}
\affiliation{Department of Astronomy and Astrophysics, Pennsylvania State University, University Park, PA 16802, USA}
\affiliation{Institute of Gravitation and the Cosmos, Center for Multimessenger Astrophysics, Pennsylvania State University, University Park,
PA 16802, USA}

\date{\today}

\begin{abstract}
We consider the prospects for future ultrahigh energy cosmic ray and neutrino observations to constrain the evolution of sources producing a proton flux above $10$~EeV (1 EeV$ = 10^{18}$~eV). We find that strong constraints on the source evolution can be obtained by combining measurements of the cosmic ray proton fraction above $30$~EeV with measurement of the neutrino flux at $1$~EeV, if neutrinos are predominantly of cosmogenic origin. In the case that interactions in the source environment produce a significant astrophysical neutrino flux, constraints on the source evolution may require measurement of the observed proton fraction, as well as, the neutrino flux at multiple energies, such as $1$~EeV and $10$~EeV. Finally, we show that fits to current UHECR data favor models which result in a $>30$~EeV proton fraction and $1$~EeV neutrino flux that could realistically be discovered by the next generation of experiments. 
\end{abstract}	

\maketitle

\section{Introduction}

\par
Over the past decade the Pierre Auger Observatory (Auger) has significantly rewritten our understanding of the cosmic ray (CR) spectrum at ultrahigh energies (UHEs). In particular, precise measurements of air shower properties have led to the conclusion that UHECRs are not predominantly protons, but that the fraction of heavier nuclei increases with energy above $10^{18.3}$~eV~\cite{Yushkov:2020nhr,PierreAuger:2014sui,
 PierreAuger:2010ymv}. 
However, there is still observational and phenomenological motivation for a flux of protons in the spectrum at the highest energies. Analysis of the distribution of depths of shower maximum, $X_\mathrm{max}$, show that the proton fraction above $10^{19}$~eV could be as high as $10\%$ in some energy bins~\cite{Bellido:2017cgf,PierreAuger:2014gko}. Furthermore, a combined  analysis of cosmic-ray composition and flux results in a non-zero proton fraction above $10^{19.5}$~eV~\cite{PierreAuger:2022atd}. Phenomenological studies have also shown that a subdominant proton component peaking above $10^{19}$~eV can significantly improve the fit to UHECR spectrum and composition data~\cite{Muzio:2019leu}.

\par
Previously it was suggested (e.g.\ \cite{vanVliet:2019nse})
that a measurement of cosmogenic
neutrinos, i.e.\ neutrinos that are produced during 
the extragalactic propagation of protons in interactions
with cosmic photon fields, can be used
to determine the cosmic ray proton fraction. But, as pointed 
out in Ref.~\cite{Moller:2018isk}, the cosmological evolution
of the sources introduces a strong degeneracy that cannot be
resolved by measurements of the neutrino flux alone. On the other hand, this implies that multimessenger studies of UHE neutrinos
and cosmic rays provide a unique opportunity to determine the 
evolution of sources. Moreover, since each candidate source class exhibits a unique redshift evolution (see Section~\ref{sec:model}), constraints
on the source evolution will provide valuable insights on the 
thus far elusive source of UHECRs.

\par
In this paper we consider the prospects for using both cosmic ray and neutrino measurements to constrain the evolution of a population of UHE proton sources, while taking into account constraints imposed by UHECR spectrum and composition, neutrino, and gamma-ray data. We show that such constraints are possible even when significant source interactions are considered. 

\section{Model}\label{sec:model}

\par
We adopt the phenomenological Unger-Farrar-Anchordoqui (UFA) CR source model~\cite{Unger:2015laa}, as elaborated in \cite{Muzio:2019leu,Muzio:2021zud}. The UFA model accounts for UHECR interactions with photons and gas in the environment surrounding the accelerator to explain the observed UHECR spectrum and composition, without assuming a particular astrophysical source type. Instead, this model uses general parameters to characterize the source's environment, such as the average number of interactions before escape and the temperature of the ambient photon field. For this study we consider the superposition of two UFA-like source populations: 1) a baseline population which accounts for the majority of the observed UHECR spectrum and composition; and 2) a population which accelerates a pure-proton spectrum to energies $\gtrsim10$ EeV. To minimize the number of free parameters we assume both populations follow the same source evolution, but a more detailed study could be done to explore the effect of a superposition of CR source populations with distinct evolutions. Additionally, in order to set conservative neutrino constraints, we assume the spectral shape of the ambient photon field to be well-characterized by a black-body spectrum for both populations~\footnote{Other parametrizations of the source photon field, including a broken power-law, were explored but did not have a significant effect on our results, in agreement with the results of \cite{Unger:2015laa,Fiorillo:2021hty}.}.

\par
We consider a two-parameter model of the source evolution $\xi(z)$, the comoving CR power density at redshift $z$ relative to its value today, consisting of a simple power law and an exponential cutoff

\begin{align}
    \xi_{m,z_0}(z) = 
    \begin{cases} 
      (1+z)^m & z \leq z_0 \\
      (1+z_0)^m e^{-(z-z_0)} & z > z_0
   \end{cases},
\end{align}

\noindent
where $-7 \leq m \leq 7$ and $1 \leq z_0 \leq 5$. This simple parametrization sufficiently captures the qualitative features of many observed source evolutions considered when modeling UHECRs and the neutrinos they produce. Additionally, several observationally-informed source evolutions are considered. These include a star formation rate (SFR) evolution~\cite{Robertson:2015uda},

\begin{align}
    \xi_\mathrm{SFR}(z) \propto \frac{(1+z)^{3.26}}{1+[(1+z)/2.59]^{5.68}},
\end{align} 

\noindent
an active galactic nuclei (AGN) evolution \cite{Stanev:2008un}, 

\begin{align}
    \xi_\mathrm{AGN}(z) \propto 
    \begin{cases} 
      (1+z)^5 & z \leq 1.7 \\
      (1+1.7)^5 & 1.7 < z \leq 2.7 \\
      (1+1.7)^5 e^{-(z-2.7)} & z > 2.7
   \end{cases},
\end{align}

\noindent
and a gamma-ray burst (GRB) evolution \cite{Kistler:2007ud,Yuksel:2008cu}, 

\begin{align}
    \xi_\mathrm{GRB}(z) \propto \frac{(1+z)^{1.5}}{\left[(1+z)^{-34} + \left(\frac{1+z}{5160}\right)^{3} + \left(\frac{1+z}{9}\right)^{35}\right]^{0.1}}.
\end{align}

\par
Throughout this work we fit the observed spectrum and composition data of Auger~\cite{PierreAuger:2020qqz,PierreAuger:2020kuy,Verzi:2020opp,Yushkov:2020nhr}, adopting a $+20\%$ shift of the Auger energy scale and a $-10$ g/cm$^2$ average shift of $\langle X_\mathrm{max} \rangle$ following~\cite{Muzio:2021zud}. The goodness-of-fit is determined by calculating a combined $\chi^2$ to the UHECR spectrum and the first two moments of the depth of shower maximum distributions, $\langle X_\mathrm{max} \rangle$ and $\sigma\left( X_\mathrm{max} \right)$ [mapped into $\langle \ln{A} \rangle$ and $\mathrm{V}(\ln{A})$, where $A$ is the CR mass number, according to the parametrization of~\cite{PierreAuger:2013xim}].

\par
The baseline model for a given source evolution is determined by fitting the UHECR spectrum above $10^{17.5}$~eV and composition data above $10^{17.8}$~eV, assuming a single-mass injection into the source environment. Once the best-fit model is determined all source parameters of the baseline model are fixed for the remainder of the analysis. The source parameters for baseline models assuming an observationally-informed source evolution can be found in Appendix~\ref{app:baselinePars}.

\par
The pure-proton population is given its own set of source parameters and an injected spectrum with a maximum rigidity in $10-1000$~EeV range. Importantly, the average number of interactions before escape is a parameter of the model, allowing our analysis to capture both the possibility of significant source interactions and of a ``naked" accelerator, free of any significant source environment. In the latter case, CRs produced by the pure-proton population only experience interactions during extragalactic propagation, as was explored in~\cite{Muzio:2019leu,vanVliet:2019nse}. 

\par
The relative contribution of the two populations is set by a parameter $f_{pp}$,

\begin{align}
    f_{pp} = \frac{\int_{E_\mathrm{ref}}^\infty E \phi_{pp} dE}{\int_{E_\mathrm{ref}}^\infty E (\phi_{pp} + \phi_\mathrm{BL}) dE},
\end{align}

\noindent
controlling the fraction of energy escaping both source populations produced by the pure-proton population, where $E_\mathrm{ref}=10^{17}$~eV, $\phi_{pp}$ is the escaping spectrum produced by the pure-proton population, and $\phi_\mathrm{BL}$ is the escaping spectrum produced by the baseline population. 

\begin{figure}[htpb!]
    \centering
    \includegraphics[width=\linewidth]{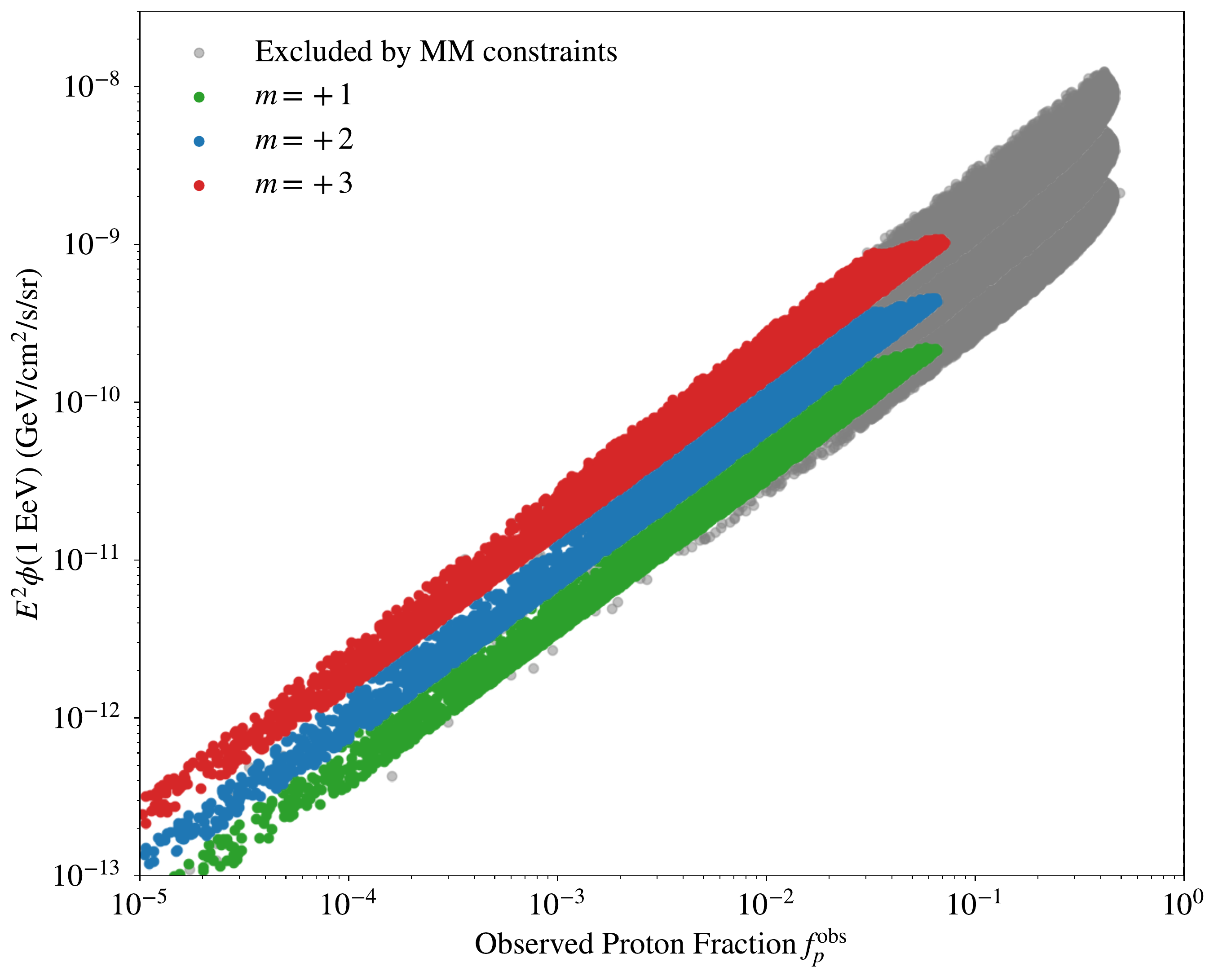}
    \caption{Correlation between the observed proton fraction above $30$~EeV, $f_p^\mathrm{obs}$, and the $1$~EeV neutrino flux, $\phi_{18}$. Each point represents a separate model realization. Gray points are model realizations excluded by multimessenger constraints. A power law source evolution was assumed for each model realization with the value of $m$ indicated by the color and $z_0 = 2$. The \textsc{Sibyll2.3c} hadronic interaction model (HIM) was assumed.}
    \label{fig:cosmo_correlation}
\end{figure}

\begin{figure}[htpb!]
    \centering
    \includegraphics[width=\linewidth]{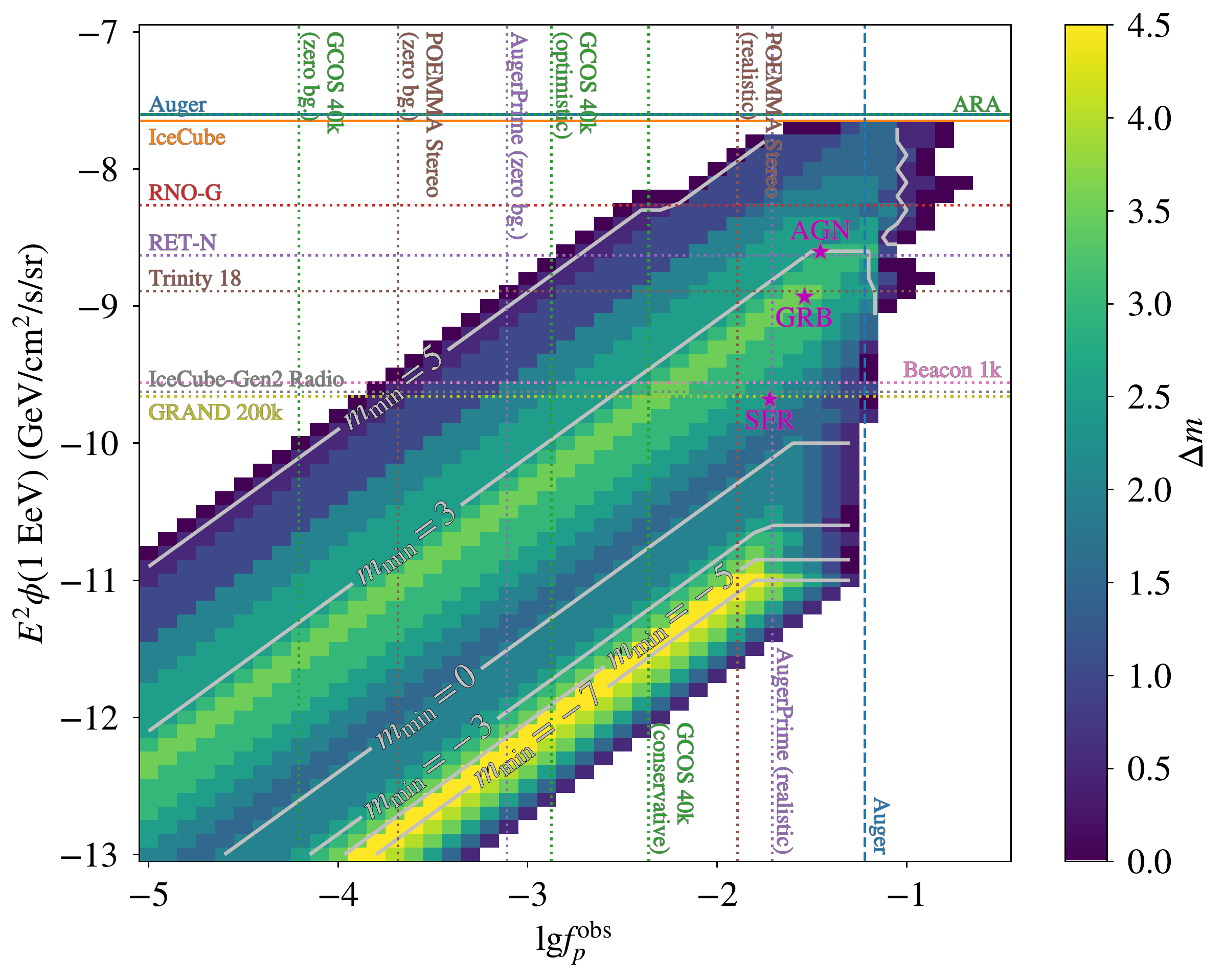}
    \caption{The range to which the power-law index of the source evolution can be constrained, $\Delta m$, for a given measurement of the observed proton fraction $f_p^\mathrm{obs}$ above $30$~EeV and $1$~EeV cosmogenic neutrino flux, marginalizing over the cutoff redshift $z_0$ and assuming \textsc{Sibyll2.3c}. Contours indicate the lower-bound on the power law index for a given measurement. The decrease in $\Delta m$ below $m_\mathrm{min}=-7$ is due to the finite range of negative values of $m$ explored. White regions indicate combinations of the observed proton fraction and neutrino flux that are either incompatible with multimessenger data, are not realizable physically, or require a source evolution with $m < -7$. Magenta stars indicate the predicted values for the best-fit models assuming SFR, AGN, and GRB source evolutions. Also indicated are current $90\%$~confidence level (CL) upper-limits on the neutrino flux for Auger~\cite{PierreAuger:2019ens} (horizontal solid light-blue line) and IceCube~\cite{IceCube:2018fhm} (horizontal solid orange line), as well as, $90\%$~CL limit forecasts from~\cite{Ackermann:2022rqc} for a variety of ongoing and future neutrino experiments~\cite{ARA:2019wcf,RNO-G:2020rmc,RNO-G:2021hfx,Prohira:2019glh,Brown:2021ane,Wissel:2020sec,Ackermann:2022rqc,GRAND:2018iaj} (horizontal dotted colored lines). The Auger measurement~\cite{Bellido:2017cgf,PierreAuger:2014sui} of the observed proton fraction above $30$~EeV is also shown for \textsc{Sibyll2.3c} (dashed vertical light blue line indicates the $1\sigma$ upper-limit). Also indicated are the $90\%$~CL limit forecasts for a variety of ongoing and future UHECR experiments~\cite{PierreAuger:2016qzd,POEMMA:2020ykm,Horandel:2021prj} (vertical dotted colored lines; details in Appendix~\ref{app:UHECRsensitivity}).}
    \label{fig:cosmo_nu18_dM_sibyll}
\end{figure}

\par
To explore the range of multimessenger signals which can be produced by the pure-proton population for each source evolution and value of $f_{pp}$, all model parameters of this population are randomly sampled, all parameters of the baseline population are held fixed~\footnote{This was done for computational efficiency, as we found that also refitting the source parameters of the baseline model had a negligible effect on our results.}, and only those controlling properties of the Galactic CR spectrum (specifically its composition, spectral index, cutoff energy, and normalization) are tuned to obtain the best-fit to the UHECR spectrum and composition data above $10^{18}$~eV. Once all parameters have been set, several criteria are used to determine whether the resulting multimessenger signals are compatible with multimessenger data. First, we require that the fit to UHECR spectrum and composition data result in a $\chi^2/\mathrm{ndf} < 5$. This cut was chosen to ensure a standard on the absolute quality of the fit, while also accommodating the varying quality of fit possible for a baseline population alone assuming different hadronic interaction models (HIMs) and source evolutions~\cite{Muzio:2019leu}. Second, pure-proton models which degrade the quality of fit by more than $3\sigma$ compared to the baseline model alone are considered to be in conflict with UHECR data~\footnote{We follow the PDG~\cite{ParticleDataGroup:2020ssz, Rosenfeld:1975fy} defining the number of sigma from the best-fit as $N_\sigma' = S^{-1} \sqrt{\chi^2_\mathrm{model} - \chi^2_\mathrm{min}}$, where $S = \sqrt{\chi^2_\mathrm{min}/N_\mathrm{dof}}$ is the scale factor introduced  to enlarge the uncertainties to account for a $\chi^2_\mathrm{min}/N_\mathrm{dof} > 1$~\cite{Rosenfeld:1975fy}, $\chi^2_\mathrm{model}$ is the $\chi^2$ for a given model, $\chi^2_\mathrm{min}$ is the $\chi^2$ of the best-fit model, and $N_\mathrm{dof}$ is the number of degrees of freedom. In order to calculate the number of sigma a model has improved (or degraded) the fit compared to the baseline model we define $N_\sigma^\mathrm{rel} = \sgn(\chi^2_\mathrm{model}-\chi^2_\mathrm{BL}) S^{-1} \sqrt{| \chi^2_\mathrm{model} - \chi^2_\mathrm{BL}|}$, where $\chi^2_\mathrm{min} = \min(\chi^2_\mathrm{model}, \chi^2_\mathrm{BL})$ in $S$.}. Third, models which produce more than $4.74$ neutrinos above $10^{15.9}$~eV are rejected at $99\%$ confidence level (CL)~\cite{Feldman:1997qc} as they violate constraints from IceCube~\cite{IceCube:2018fhm,IceCube:2021rpz}. Finally, we consider limits on the gamma-ray flux at GeV--TeV energies, from \textit{Fermi}-LAT~\cite{Fermi-LAT:2014ryh,Fermi-LAT:2015otn}, and at EeV energies from Auger~\cite{Rautenberg:2021vvt,PierreAuger:2021mjh,PierreAuger:2022uwd}, but find that no models compatible with other multimessenger constraints are capable of violating them. This combination of constraints limits $m\leq + 6$.

\par
It is possible that some realizations of this model will have parameters which imply a large source with a strong magnetic field. In this case, pions and muons produced in the environment suffer significant synchrotron losses before decaying, effectively cutting-off the resulting neutrino spectrum. We find that excluding such model realizations does not change our results. However if, in reality, UHECR environments are in a regime where synchrotron losses are significant, only the results of Section~\ref{sec:cosmogenic} would be applicable.

\section{Results}

\subsection{Cosmogenic-only case} \label{sec:cosmogenic}

\par
For a fixed source evolution we find the flux of neutrinos at $1$~EeV, $\phi_{18}$, to have a strong correlation with the observed proton fraction above $30$~EeV, $f^\mathrm{obs}_p$, so that $\phi_{18} \propto f^\mathrm{obs}_p$ (see Fig.~\ref{fig:cosmo_correlation}), as was reported by~\cite{vanVliet:2019nse}. To capture the dispersion of this correlation we find the maximum and minimum values of the observed proton fraction-to-$1$~EeV neutrino flux ratio, $r_{p\nu,18}$, among all models compatible with multimessenger data. By construction, all models must then obey $r^\mathrm{min}_{p\nu,18} \leq r_{p\nu,18} \leq r^\mathrm{max}_{p\nu,18}$. For a fixed source evolution, this fact allows a constraint to be placed on either $f^\mathrm{obs}_p$ or $\phi_{18}$ if the other quantity is known according to:

\begin{align}
    f^\mathrm{obs,min}_p &= \min\left(r_{p\nu,18}^\mathrm{min} \tilde{\phi}_{18}, F_p^\mathrm{obs, max}\right)~, \label{eq:fpmin}\\
    f^\mathrm{obs,max}_p &= \min\left(r_{p\nu,18}^\mathrm{max} \tilde{\phi}_{18}, F_p^\mathrm{obs, max}\right)~, \label{eq:fpmax}\\
    \phi_{18}^\mathrm{min} &= \min\left(\frac{\tilde{f_p}^\mathrm{obs}}{r_{p\nu,18}^\mathrm{max}}, \Phi^\mathrm{max}_{18}\right)~, \label{eq:phi18min}\\
    \phi_{18}^\mathrm{max} &= \min\left(\frac{\tilde{f_p}^\mathrm{obs}}{r_{p\nu,18}^\mathrm{min}}, \Phi^\mathrm{max}_{18}\right)~ \label{eq:phi18max},\\
    \intertext{where}    
    \tilde{f_p}^\mathrm{obs} &= \min\left(f_p^\mathrm{obs}, F_p^\mathrm{obs, max}\right) \\
    \intertext{and}   
    \tilde{\phi}_{18} &= \min\left(\phi_{18}, \Phi^\mathrm{max}_{18}\right)
\end{align}

\noindent
are the observed proton fraction and $1$~EeV neutrino flux truncated at their maximum realizable values compatible with multimessenger data, $F_p^\mathrm{obs, max}$ and $\Phi^\mathrm{max}_{18}$ ~\footnote{A similar analysis could be carried out by considering the integral UHE photon flux above $10$~EeV, as this observable also correlates well with both the observed proton fraction and the neutrino flux. However, due to the short attenuation length of UHE photons, the results are very sensitive to the assumed distance to the nearest source. For this reason we leave such an analysis for future work.}. 

\par
To be conservative we subtract off the baseline source population's contribution to $f_p^\mathrm{obs}$ and $\phi_{18}$, but in principle either of these values may have a ``floor'' set by the baseline source population depending on the true evolution of these sources. With these constraints in hand one can determine the range of source evolutions 

\begin{align}
    \Delta m = m_\mathrm{max} - m_\mathrm{min}
\end{align}

\noindent
which satisfy ${f_p^\mathrm{obs,min} \leq f_p^\mathrm{obs} \leq f_p^\mathrm{obs, max}}$ and ${\phi_{18}^\mathrm{min} \leq \phi_{18} \leq \phi_{18}^\mathrm{max}}$ for a particular $(f_p^\mathrm{obs}, \phi_{18})$. These constraints are shown in Fig.~\ref{fig:cosmo_nu18_dM_sibyll} assuming the \textsc{Sibyll2.3c}~\cite{Fedynitch:2018cbl} HIM. Analogous figures showing the results when assuming the \textsc{EPOS-LHC}~\cite{Pierog:2013ria} HIM can be found in Appendix~\ref{app:eposResults}. 

\par
It is clear that for the cosmogenic-only case, the dispersion of the correlation between the observed proton fraction and $1$~EeV neutrino flux is small (as evidenced by Fig.~\ref{fig:cosmo_correlation} and the small value of $\Delta m$ over most of the parameter space in Fig.~\ref{fig:cosmo_nu18_dM_sibyll}). This enables UHECR and neutrino measurements to jointly measure the evolution of such a population of UHE proton sources. For example, Fig.~\ref{fig:cosmo_nu18_dM_sibyll} shows if AugerPrime~\cite{PierreAuger:2016qzd} measures the proton fraction above $30$~EeV to be $3\%$, then assuming the Radio Echo Telescope for Neutrinos (RET\=/N)~\cite{Prohira:2019glh} measures the $1$~EeV neutrino flux to be $10^{-8.5}$~GeV$/$cm$^2/$s$/$sr, the source evolution will be constrained $\Delta m \lesssim 3$ and $m > 3$. 

\begin{figure*}[htbp!]
    \centering
	\begin{minipage}{0.49\linewidth}
	  \centering
      \subfloat[\label{fig:nu18_dM_sibyll}]{\includegraphics[width=\textwidth]{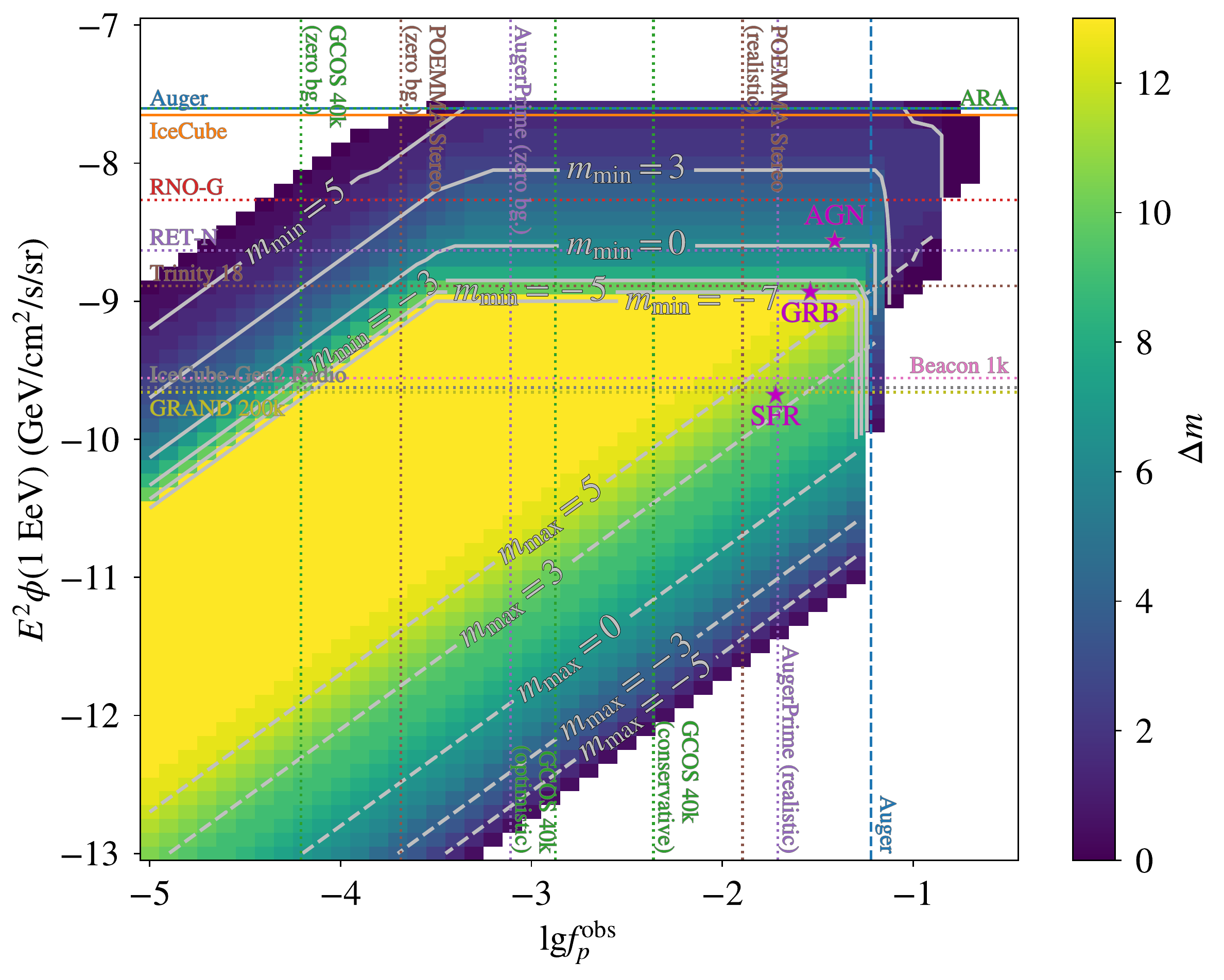}}
    \end{minipage}
    \begin{minipage}{0.49\linewidth}
	  \centering
      \subfloat[\label{fig:nu19_dM_sibyll}]{\includegraphics[width=\textwidth]{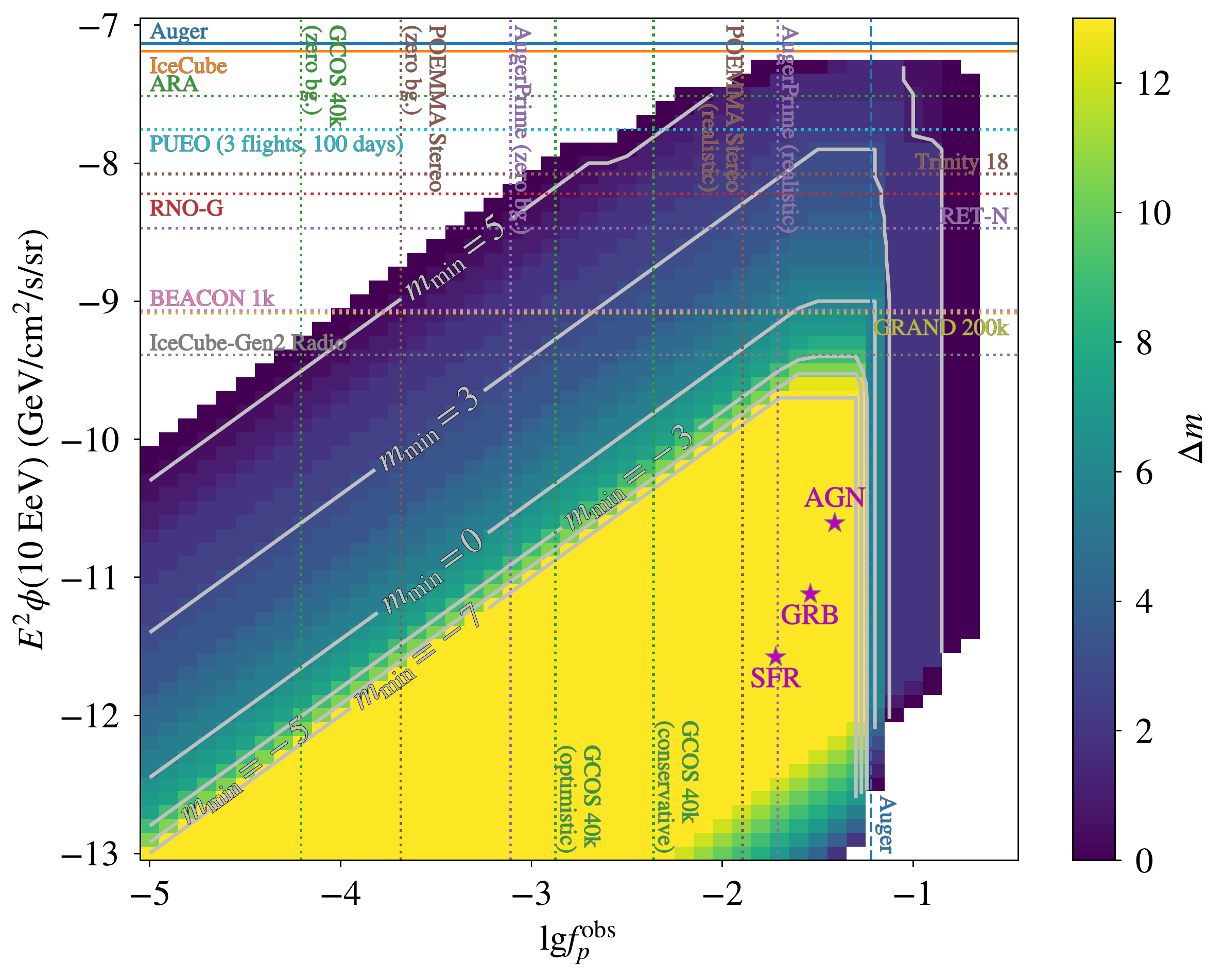}}
    \end{minipage}
	\caption{Same as Fig.~\ref{fig:cosmo_nu18_dM_sibyll} but for the general case where protons may or may not have significant interactions in the source environment, and for measuring the neutrino flux at $1$~EeV (left) and $10$~EeV (right). Dashed contours indicating the upper-bound on the source evolution's power law index are also shown for the $1$~EeV case. Additional $90\%$~CL limit forecasts for $10$~EeV neutrino sensitivity from~\cite{Ackermann:2022rqc} are shown in the top panels for a variety of ongoing and future neutrino experiments~\cite{ARA:2019wcf,RNO-G:2020rmc,RNO-G:2021hfx,Prohira:2019glh,Brown:2021ane,Wissel:2020sec,Ackermann:2022rqc,GRAND:2018iaj,PUEO:2020bnn} (dotted colored lines).}
	\label{fig:general_dM_sibyll}
\end{figure*}

\par
Fig.~\ref{fig:cosmo_nu18_dM_sibyll} also shows that a $1$~EeV neutrino detection alone will be able to constrain the source evolution of such UHE proton sources. For example, Fig.~\ref{fig:cosmo_nu18_dM_sibyll} shows that if the Radio Neutrino Observatory in Greenland (RNO\=/G)~\cite{RNO-G:2020rmc,RNO-G:2021hfx} detects a $1$~EeV neutrino then source evolutions with $m<3$ will be excluded. This is simply because these source evolutions are not capable of producing a large enough $1$~EeV neutrino flux to be detectable by RNO\=/G without violating current CR composition constraints. 

\par
Finally, we comment that Fig.~\ref{fig:cosmo_nu18_dM_sibyll} can be used to provide benchmark sensitivities for future UHECR and neutrino experiments. Measurement, or constraint, on the flux of one messenger places an upper-bound on the flux of the other messenger. For example, Fig.~\ref{fig:cosmo_nu18_dM_sibyll} shows if the observed proton fraction is constrained to be less than $1\%$ then that would imply the $1$~EeV neutrino flux is less than $10^{-8}$~GeV$/$cm$^2/$s$/$sr. A similar statement for the observed proton fraction is possible under very mild assumptions about the source evolution. For example, the $1$~EeV neutrino flux were constrained to be less than $10^{-10}$~GeV$/$cm$^2/$s$/$sr then the observed proton fraction must be less than $5\%$ for positive source evolutions.

\subsection{General case} \label{sec:general}

\par
In the more general case, where we allow for the possibility of a significant number of interactions in the environment host to the accelerator, the correlation between the flux of neutrinos at $1$~EeV and the observed proton fraction above $30$~EeV is weaker. This weaker correlation amounts to a wider dispersion and, therefore, a larger range of $r_{p\nu,18}$ values. The resulting constraints for this more general case are shown in Fig.~\ref{fig:nu18_dM_sibyll}. In particular, whereas in the cosmogenic-only case one can always effectively measure the source evolution, in the general case this is only possible for certain $(f_p^\mathrm{obs}, \phi_{18})$ combinations. More generally, it may only be possible to set an upper- or lower-bound on the value of $m$ using the $1$~EeV neutrino flux.  

\par
However, several planned and proposed neutrino experiments in the near future will have peak sensitivity in the $10$~EeV range, rather than the $1$~EeV range. Given the large dispersion of the $\phi_{18}$--$f^\mathrm{obs}_p$ correlation in the general case, it is worthwhile to explore how these higher energy neutrino observatories will be able to provide insight into the evolution of UHECR sources. Therefore, we also consider the correlation between the $10$~EeV neutrino flux, $\phi_{19}$, and the observed proton fraction. Similar to the $1$~EeV case, we find that $f^\mathrm{obs}_p \propto \phi_{19}$ but with a large dispersion. Defining the observed proton fraction-to-$10$~EeV neutrino flux ratio, $r_{p\nu, 19}$, we can constrain the realizable range of $f^\mathrm{obs}_p$ and $\phi_{19}$ compatible with multimessenger data analogously to equations~\eqref{eq:fpmin}-\eqref{eq:phi18max}. The constraints based on the $10$~EeV neutrino flux can be found in Fig.~\ref{fig:nu19_dM_sibyll}.

\par
From Fig.~\ref{fig:nu19_dM_sibyll} we see that it is only possible to constrain the source evolution for some combinations of $f_p^\mathrm{obs}$ and $\phi_{19}$ -- similar to the situation in Fig.~\ref{fig:nu18_dM_sibyll}. However even in the general case, where UHE protons have a significant number of interactions in the source environment, measurement of the source evolution may be possible by combining the observed proton fraction with measurement of the $1$~EeV and $10$~EeV neutrino fluxes. For example, let's assume the observed proton fraction was measured to be $1\%$, the $1$~EeV neutrino flux were measured to be $10^{-10}$~GeV$/$cm$^2/$s$/$sr, and the $10$~EeV neutrino flux were measured to be $10^{-9.5}$~GeV$/$cm$^2/$s$/$sr. Then from Fig.~\ref{fig:general_dM_sibyll} we see that these measurements would allow us to infer that $m\lesssim 3$ (driven by the $1$~EeV neutrino flux measurement) and $m\gtrsim 0$ (driven by the $10$~EeV neutrino flux measurement). However, it is important to note that not all combinations of these observables yield strong constraints on the source evolution.

\par
Tantalizingly, both Figs.~\ref{fig:cosmo_nu18_dM_sibyll} and~\ref{fig:general_dM_sibyll} show that the best-fit models assuming astrophysical source evolutions predict an observed proton fraction of $\gtrsim 1\%$ regardless of the particular astrophysical scenario. These best-fit models also predict a $1$~EeV neutrino flux that will be detectable by the next generation of neutrino experiments. This prediction suggests that discovery of such a UHE proton component and its neutrino flux -- and therefore a measurement of the evolution of its sources -- by the next generation of UHECR and neutrino experiments is a realistic possibility.

\section{Summary}

\par
In this study we have considered the prospects for constraining the evolution of a population of pure-proton sources by combining UHECR and neutrino data. Neither of these messengers can determine the source evolution alone. However, we have found that near-future UHECR and neutrino detectors could realistically place strong constraints on the evolution of such a population.

\par
In the case of a purely cosmogenic flux of neutrinos, near-future detectors will constrain the source evolution as long as the proton fraction above $30$~EeV is $\gtrsim10^{-4}$ and the neutrino flux at $1$~EeV is $\gtrsim 10^{-10}$ GeV/cm$^2$/s/sr --  a requirement favored by the best-fits to the UHECR spectrum and composition data we find. In this case, the $1$~EeV neutrino flux and the observed proton fraction can be combined to constrain the source evolution to a narrow range of possibilities. 

\par
In the case that source interactions result in a significant astrophysical neutrino flux, more information may be required to constrain the source evolution. We have shown that by combining the observed proton fraction with measurements of the $1$~EeV and $10$~EeV neutrino flux, future detectors may be able to constrain the source evolution's power-law index $m$ to a limited interval. Even if the neutrino flux is only measured at one of these energies, an upper- or lower-bound may still be placed on $m$.

\par
Importantly, even if source interactions are significant, best-fit models still predict that this proton component and its secondary neutrinos at $1$~EeV will be detectable by the next generation of UHECR and neutrino experiments for many of the evolutions often considered for the sources of UHECRs. 

\par
Our results underscore the complementarity of neutrino and UHECR detectors, as well as, the need for a next-generation of detectors for both of these messengers. Perhaps epitomizing the strength of multimessenger astrophysics, our results show how combining neutrino and UHECR observations provides access to a quantity inaccessible by either of these messengers alone. \\

\acknowledgments

 We thank Foteini Oikonomou, David Seckel, and Domenik Ehlert for useful feedback on our analysis. The research of MSM is supported by the NSF MPS-Ascend Postdoctoral Award \#2138121. The research of SW is supported by NSF Awards \#2111232 and \#2033500 and NASA grants 80NSSC20K0925, 80NSSC22K1519 and 80NSSC21M0116.

\bibliography{main}


\appendix 

\section{Estimation of UHECR experimental sensitivity to $f^\mathrm{obs}_p$}\label{app:UHECRsensitivity}

\par
To estimate the ability of ongoing and future UHECR experiments to constrain $f^\mathrm{obs}_p$ we consider two possible cases. First, we consider the optimistic case where measurement of the proton flux is background-free (i.e. the experiment can perfectly separate protons from observed heavier nuclei). In this case, the strongest limit would be set if no proton events are detected, then the $90\%$ CL upper-limit is given by

\begin{align}
    f^\mathrm{obs,\text{zero bg.}}_{p} = \frac{\mathrm{FC}(0,0)}{N_\mathrm{evts}}~,
\end{align}

\noindent
where $\mathrm{FC}(0,0) = 2.44$ is $90\%$~CL Feldman-Cousins upper-limit for zero observed and background events, and $N_\mathrm{evts}$ corresponds to the total number of CR events detected above $30$~EeV. 

\par
\begin{figure}[htpb!]
    \centering
    \includegraphics[width=\linewidth]{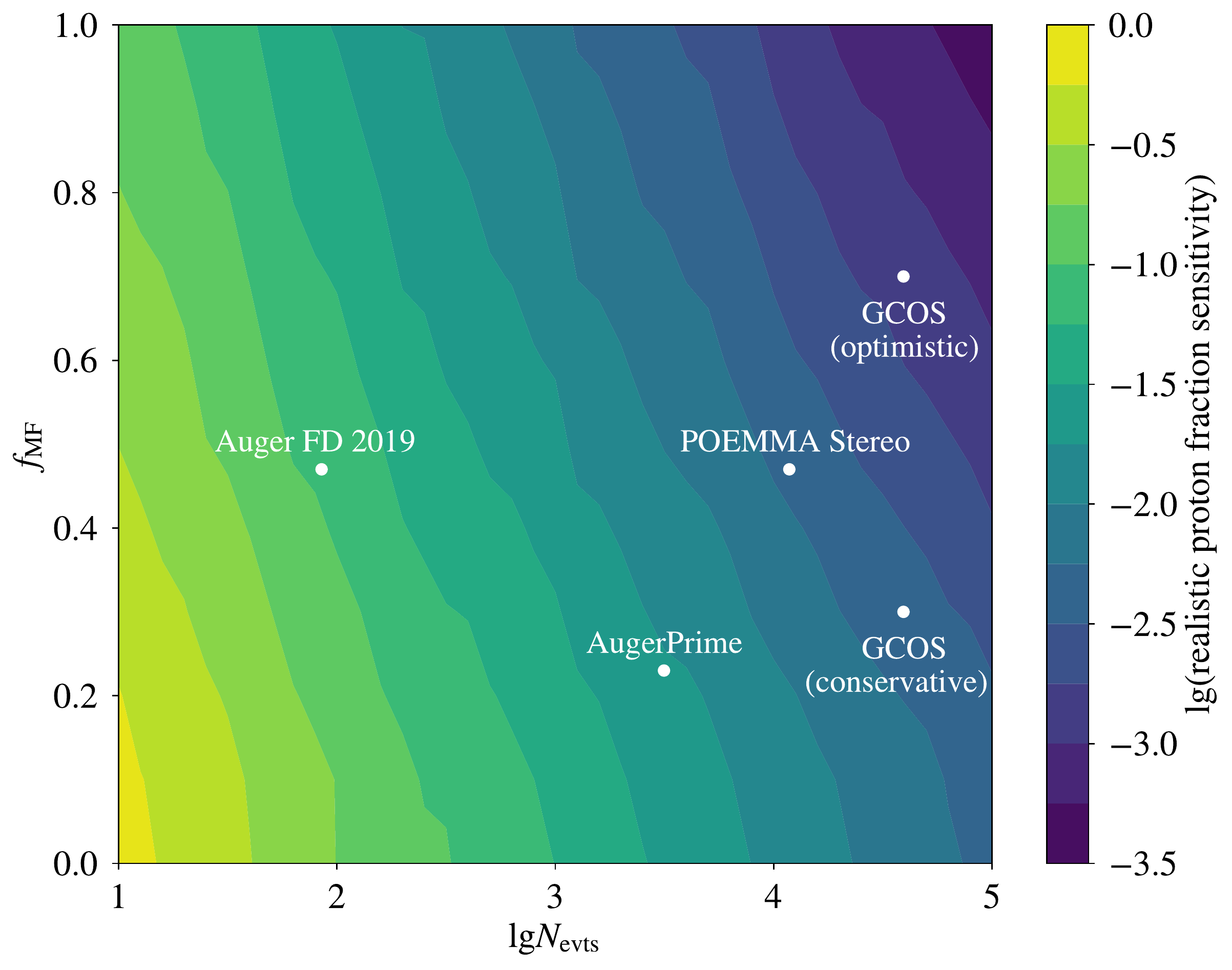}
    \caption{Estimate of the $90\%$~CL 
    proton sensitivity in the presence of helium background above $3\times 10^{19}$~eV for a given proton-helium merit factor, $f_\mathrm{MF}$, and observed number of events, $N_\mathrm{evts}$. Further details can be found in the text. The values of $f_\mathrm{MF}$ and $N_\mathrm{evts}$ assumed for each of the experiments considered are also indicated.}
    \label{fig:protonSensi}
\end{figure}

More generally, we consider the case where the proton flux cannot be perfectly separated from heavier nuclei, but a considerable overlap exists in the distributions of the mass-sensitive experimental variable $Y$ (e.g.\ the shower maximum  $X_\mathrm{max}$). Conservatively, we assume a background of helium events, since this is the nucleus which is most difficult to separate from protons. We define the proton-fraction sensitivity
as the minimum fraction with which the null-hypothesis (pure helium flux) can be rejected at a confidence level of $90\%$~CL. This minimum fraction is determined by repeatedly sampling $Y$ distributions of $N_\mathrm{evts}$ helium events. To each of these simulated data sets we fit a two-component (proton and helium) model. The 90\% quantile of the obtained proton fraction distribution for a pure helium composition defines then the sensitivity for the proton fraction. Figure~\ref{fig:protonSensi} shows the resulting sensitivity as a function of these two variables assuming $Y$ is normally distributed~\footnote{For $Y=X_{\rm max}$ we compared the sensitivities for two-component normal and Gumbel distributions with the same merit factor and the same ratio of standard deviations, $\sigma(Y_\mathrm{He})/\sigma(Y_p)$. The resulting sensitivities are of similar magnitude, about 40\% better for the normal distribution than for the skewed Gumbel distribution for proton and helium primaries.}.

\par
For this purpose, we obtain the expected number of events above $30$~EeV for each experiment by multiplying its expected exposure by the integral CR flux above $30$~EeV according to the Auger spectrum model~\cite{PierreAuger:2021ibw}. Experimental exposures, $\mathcal{E}$, were taken from~\cite{Coleman:2022abf}.

\par
Our assumptions about the mass sensitive variable $Y$ vary depending on the experiment. For POEMMA we use $X_\mathrm{max}$ as the mass-sensitive variable $Y$, distributed according to a generalized Gumbel distribution with parameter values from~\cite{Arbeletche:2019qma}. For AugerPrime and GCOS we assume that $\sigma(Y_\mathrm{He})/\sigma(Y_p) = \sigma(X_\mathrm{max,He})/\sigma(X_{\mathrm{max,}p}) \simeq 0.71$ to determine the separation of the proton and helium distributions for a given merit factor. For AugerPrime we use the published proton-helium merit factors~\cite{PierreAuger:2016qzd}, while for GCOS we consider a high- and low-resolution design with proton-helium merit factors of $0.7$~(optimistic) and $0.3$~(conservative), respectively. In both cases we assume $Y$ follows a normal distribution.

\section{Results for \textsc{EPOS-LHC}}\label{app:eposResults}

\begin{figure}[htpb!]
    \centering
    \includegraphics[width=\linewidth]{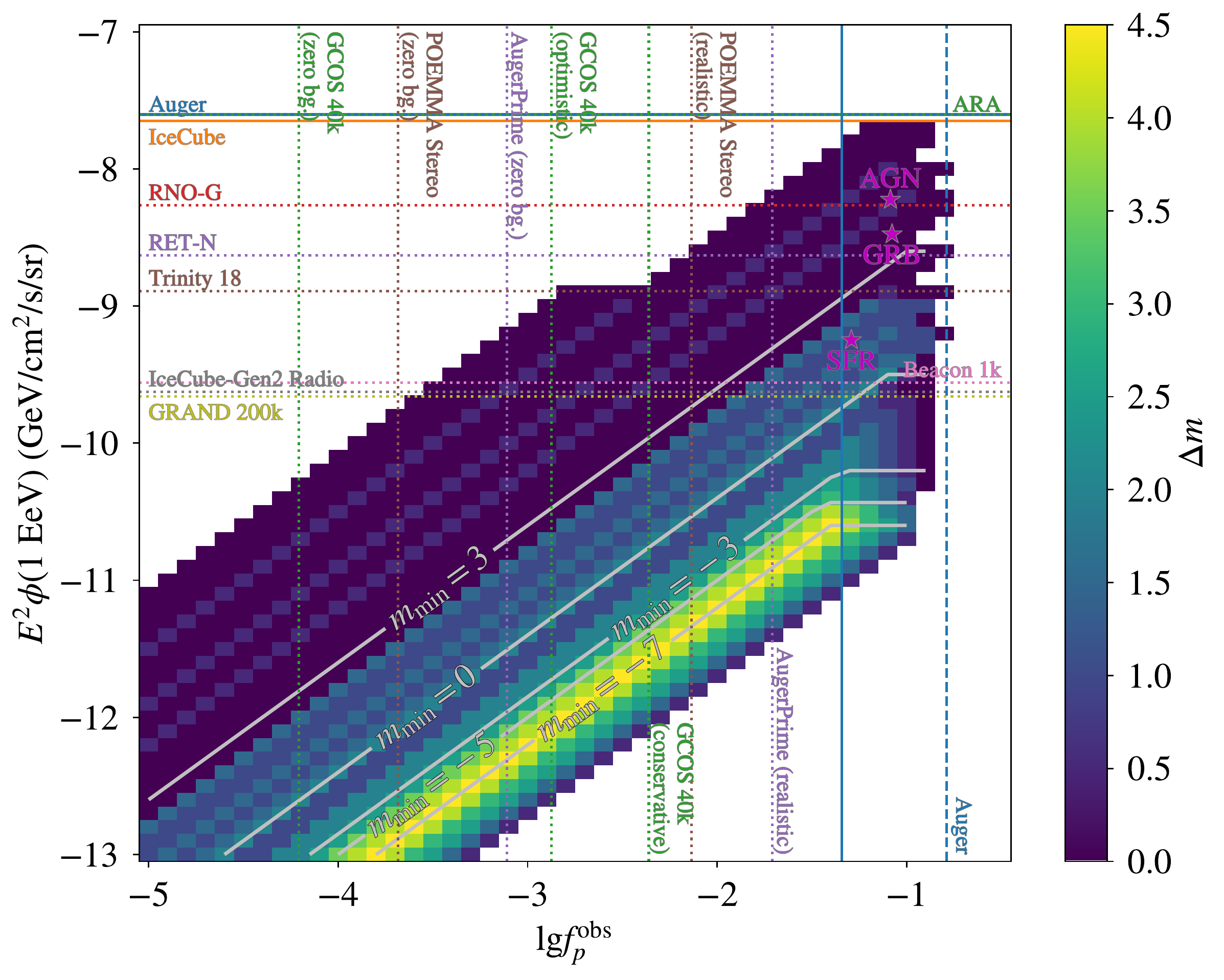}
    \caption{Same as Fig.~\ref{fig:cosmo_nu18_dM_sibyll} but using \textsc{EPOS-LHC} as the HIM. The Auger measurement~\cite{Bellido:2017cgf,PierreAuger:2014sui} of the observed proton fraction above $30$~EeV is shown for \textsc{EPOS-LHC} (vertical light blue line, central value is solid and $1\sigma$ errors are dashed).}
    \label{fig:cosmo_nu18_dM_epos}
\end{figure}

\par
Figure~\ref{fig:cosmo_nu18_dM_epos} shows the range of power-law indices allowed for a particular combination of the observed proton fraction and the $1$~EeV neutrino flux for the cosmogenic-only case, when assuming the \textsc{EPOS-LHC} HIM. Compared with Fig.~\ref{fig:cosmo_nu18_dM_sibyll}, \textsc{EPOS-LHC} generally allows for a larger proton fraction, due to the fact that it infers the composition to be lighter from air shower data. This difference in interpretation of air shower data also leads Auger to infer a non-zero proton fraction of $\sim 5\%$~\cite{Bellido:2017cgf}. The best-fit models for astrophysically-informed source evolutions also favor larger proton fractions at Earth.

\par
The most noticeable difference between Fig.~\ref{fig:cosmo_nu18_dM_sibyll} and Fig.~\ref{fig:cosmo_nu18_dM_epos}, though, is the much smaller range of $\Delta m$ values in the \textsc{EPOS-LHC} case. This is due to the narrower dispersion in the $f_p^\mathrm{obs}-\phi_{18}$ correlation, driven by the fact that fits to UHECR data assuming \textsc{EPOS-LHC} generally have a poorer quality and that we require $\chi^2/ndf < 5$. This results in models assuming \textsc{EPOS-LHC} effectively being more constrained than those assuming \textsc{Sibyll2.3c}.

\par
\begin{figure*}[htpb!]
    \centering
    \begin{minipage}{0.49\linewidth}
	  \centering
      \subfloat[\label{fig:nu18_dM_epos}]{\includegraphics[width=\textwidth]{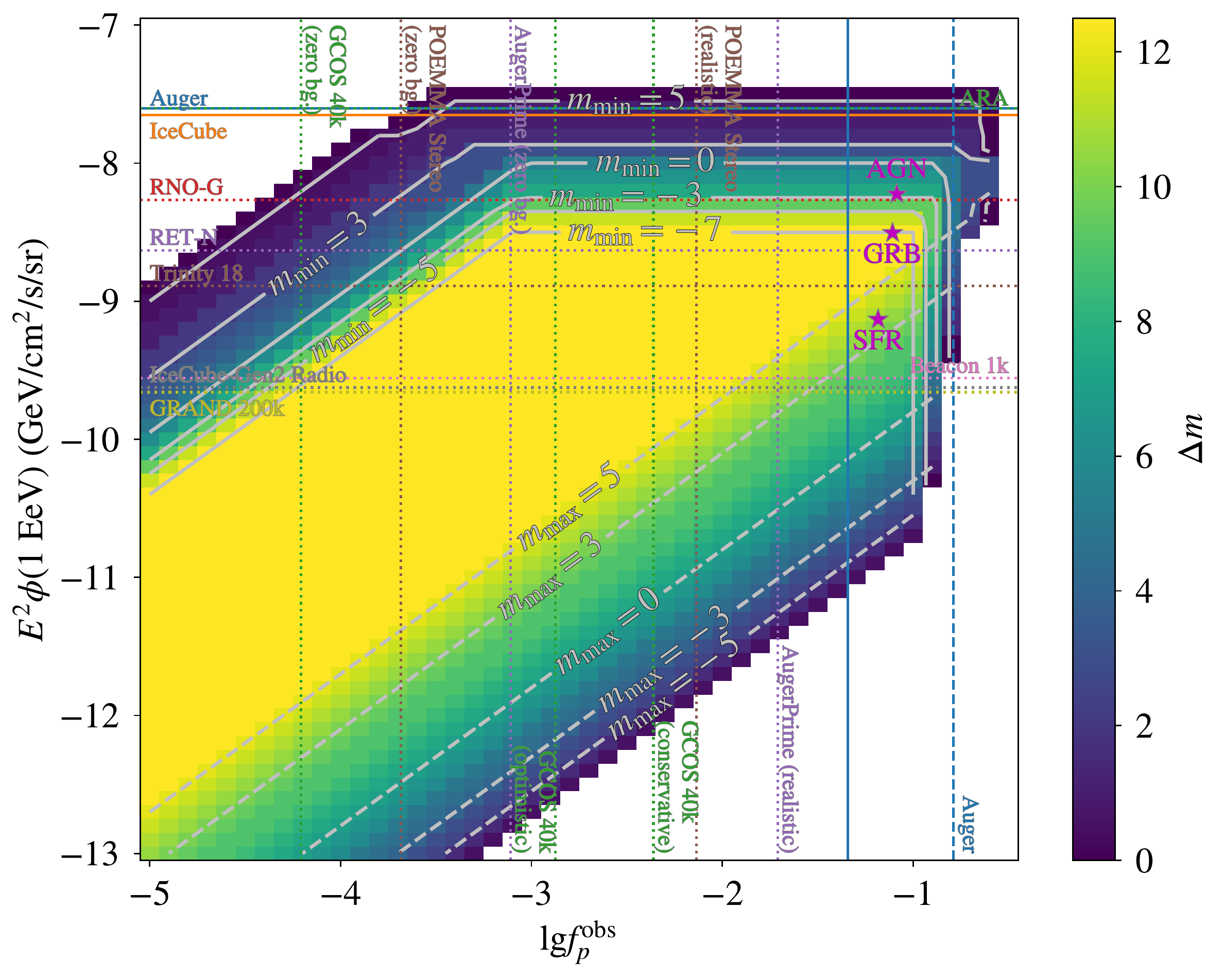}}
    \end{minipage}
    \begin{minipage}{0.49\linewidth}
	  \centering
      \subfloat[\label{fig:nu19_dM_epos}]{\includegraphics[width=\textwidth]{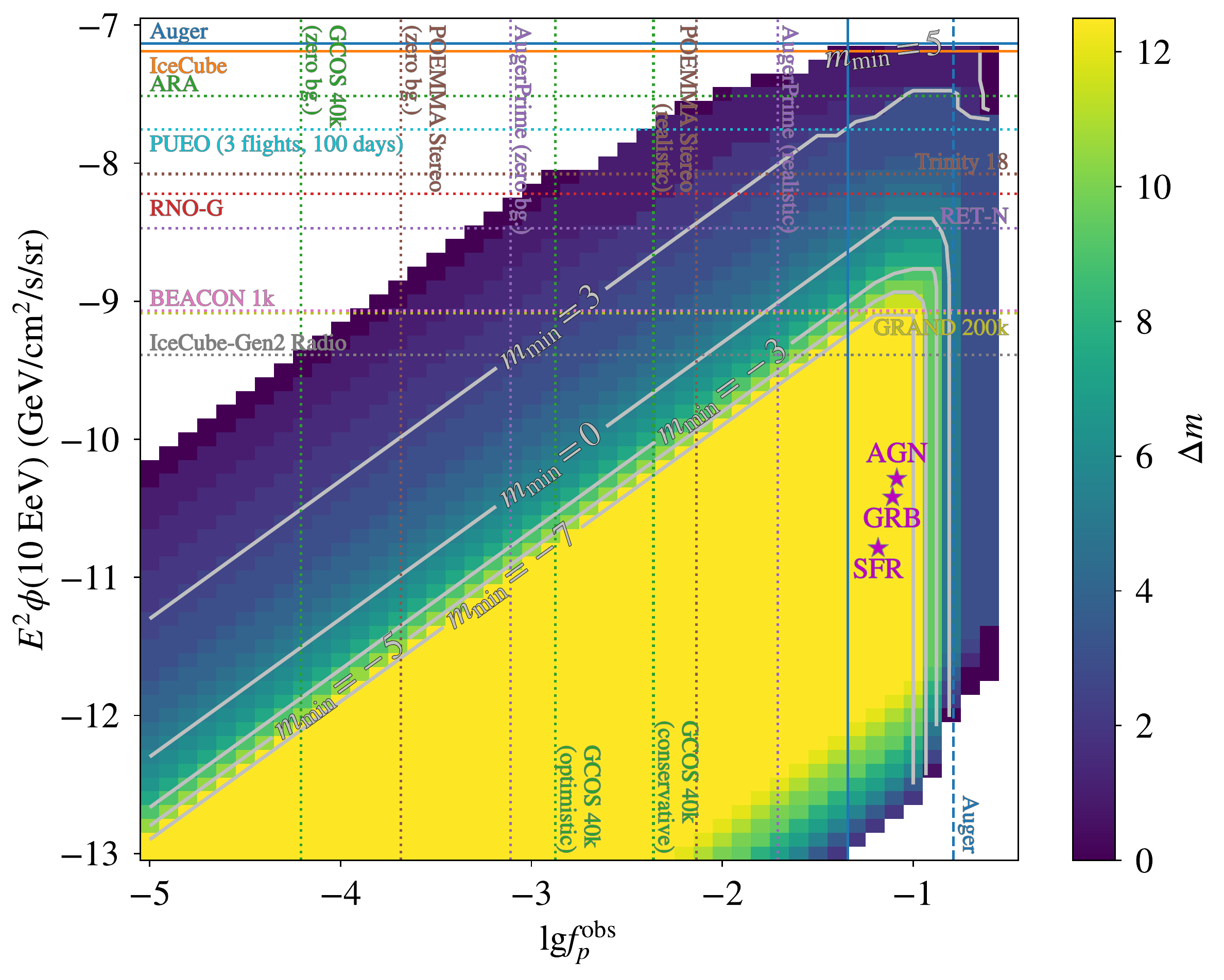}}
    \end{minipage}
	\caption{Same as Fig.~\ref{fig:general_dM_sibyll} but using \textsc{EPOS-LHC} as the HIM. The Auger measurement~\cite{Bellido:2017cgf,PierreAuger:2014sui} of the observed proton fraction above $30$~EeV is shown for \textsc{EPOS-LHC} (vertical light blue line, central value is solid and $1\sigma$ errors are dashed).}
	\label{fig:general_dM_epos}
\end{figure*}

Figure~\ref{fig:general_dM_epos} shows the range of power-law indices allowed for particular combinations of the observed proton fraction with the $1$~EeV and $10$~EeV neutrino flux for the general case, where protons may or may not have significant interactions in the source environment, for \textsc{EPOS-LHC}. Similar to the general case under \textsc{Sibyll2.3c} (see Fig.~\ref{fig:general_dM_sibyll}) not all combinations of these observables lead to constraints on the source evolution, due to the large dispersion in $r_{p\nu, 18}$ and $r_{p\nu,19}$. However, for some combinations of these observables it is possible to place an upper- or lower-bound on $m$, and combining measurements of all three observables can result in strong constraints on the source evolution in some cases. 

\section{Best-fit baseline model parameters}\label{app:baselinePars}

Table~\ref{tab:baselinePars} shows the best-fit source parameters for baseline models assuming either a SFR, GRB, or AGN source evolution. The parameters in this table are as follows: $\gamma_\mathrm{inj}$ the spectral index ($J \propto E^{\gamma_\mathrm{inj}}$) at injection into the source environment; $R_\mathrm{max}$ is the maximum rigidity of the injected CR spectrum, where the spectrum is cutoff exponentially; $r_\mathrm{esc}$ is the ratio of the escape and totalinteraction times for a $10^{19}$~eV iron nucleus; $r_{g\gamma}$ is the ratio of the hadronic and photohadronic interaction times for a $10^{19}$~eV iron nucleus; $R_\mathrm{diff}$ is the characteristic rigidity scale of diffusion in the source's turbulent magnetic field; $r_\mathrm{size}$ is the ratio of the source's size to the magnetic field's coherence length~\footnote{A non-finite value of $r_\mathrm{size}$ indicates that the highest energy CRs considered in our model enter the quasi-ballistic diffusion regime before escaping the source}; $T$ is the black-body temperature of the ambient photon field surrounding the source; and, $A_\mathrm{inj}$ is the mass number of the CRs injected into the source environment (non-integers represent the average mass due to a mixture of two consecutive mass numbers in order for $A_\mathrm{inj}$ to be a continuous model parameter).

\begin{table*}
\centering
\setlength\tabcolsep{12pt}

\begin{tabularx}{\linewidth}{l S S S S S S}

\hline \hline
 & \multicolumn{2}{c}{\textbf{SFR}} & \multicolumn{2}{c}{\textbf{GRB}} & \multicolumn{2}{c}{\textbf{AGN}} \\
\textbf{Parameter} & \textsc{Sibyll2.3c} & \textsc{EPOS-LHC} & \textsc{Sibyll2.3c} & \textsc{EPOS-LHC} & \textsc{Sibyll2.3c} & \textsc{EPOS-LHC} \\
\hline
$\gamma_\mathrm{inj}$ & -1.14 & -1.71 & -1.1 & -0.0 & -0.99 & -0.55 \\
$\log_{10}(R_\mathrm{max}/\mathrm{V})$ & 18.58 & 18.76 & 18.65 & 18.48 & 18.64 & 18.58 \\
$\log_{10}{r_\mathrm{esc}}$ & 2.38 & 3.11 & 2.64 & 1.89 & 2.51 & 2.33 \\
$\log_{10}{r_{g\gamma}}$ & 9.99 & 1.02 & 9.84 & 1.26 & 6.61 & 1.7 \\
$\log_{10}(R_\mathrm{diff}/\mathrm{V})$ & 17.66 & 14.0 & 14.0 & 14.15 & 14.0 & 14.01 \\
$\tanh(\log_{10}{r_\mathrm{size}})$ & 0.8 & 1.0 & 1.0 & 1.0 & 1.0 & 1.0 \\
$T/\mathrm{K}$ & 1800 & 1003 & 5000 & 6013 & 5002 & 4007 \\
$A_\mathrm{inj}$ & 32.52 & 26.26 & 32.0 & 25.53 & 32.0 & 27.75 \\
\hline

\end{tabularx}
\caption{\label{tab:baselinePars}Best-fit source parameters for baseline models assuming an observationally-informed source evolution. Definitions of the source parameters are given in the text.}
\end{table*}

\section{Maximum UHE neutrino flux}

\par
\begin{figure*}[htpb!]
    \centering
    \begin{minipage}{0.49\linewidth}
	  \centering
      \subfloat[\label{fig:maxNu_sibyll}]{\includegraphics[width=\textwidth]{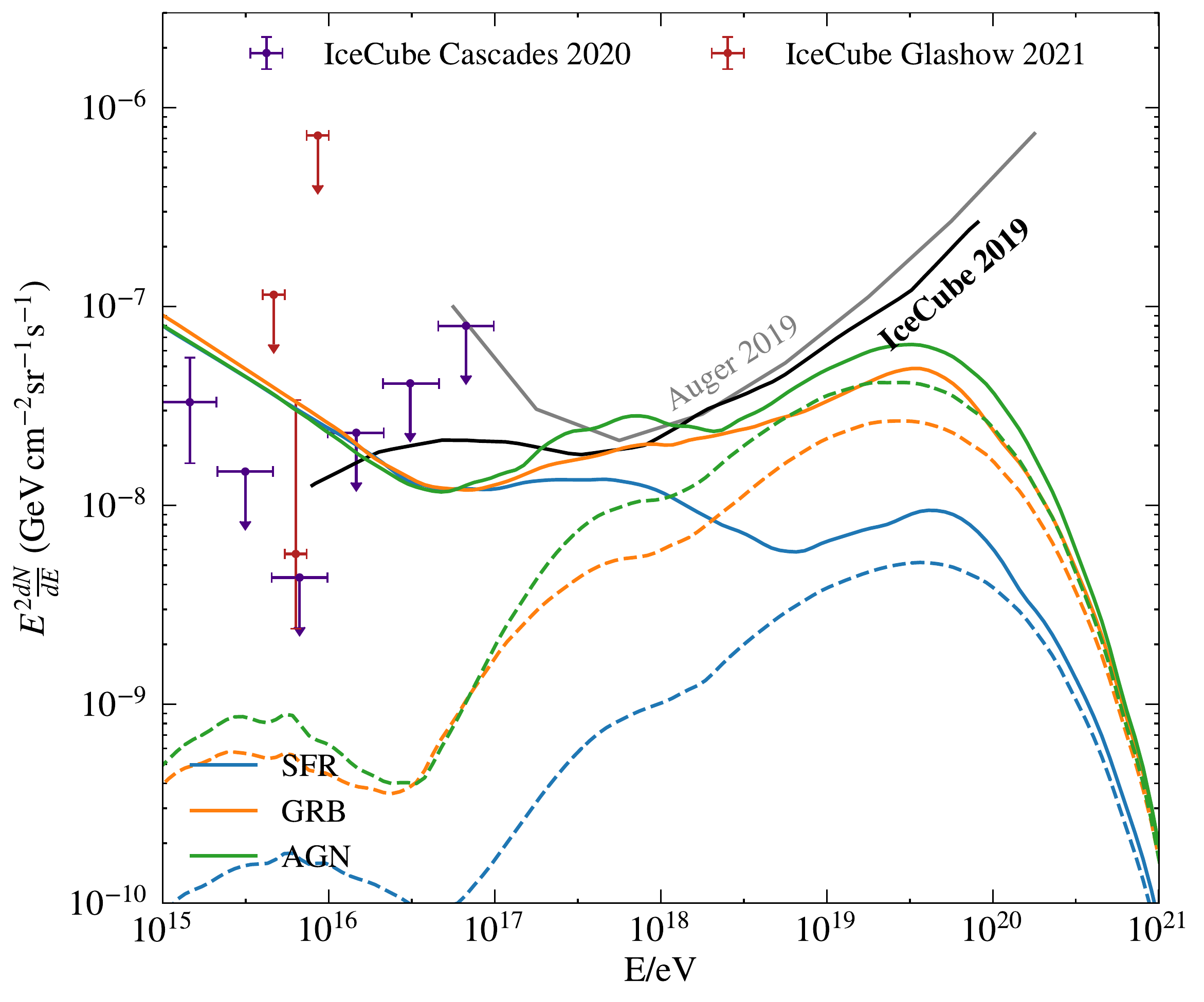}}
    \end{minipage}
    \begin{minipage}{0.49\linewidth}
	  \centering
      \subfloat[\label{fig:maxNu_epos}]{\includegraphics[width=\textwidth]{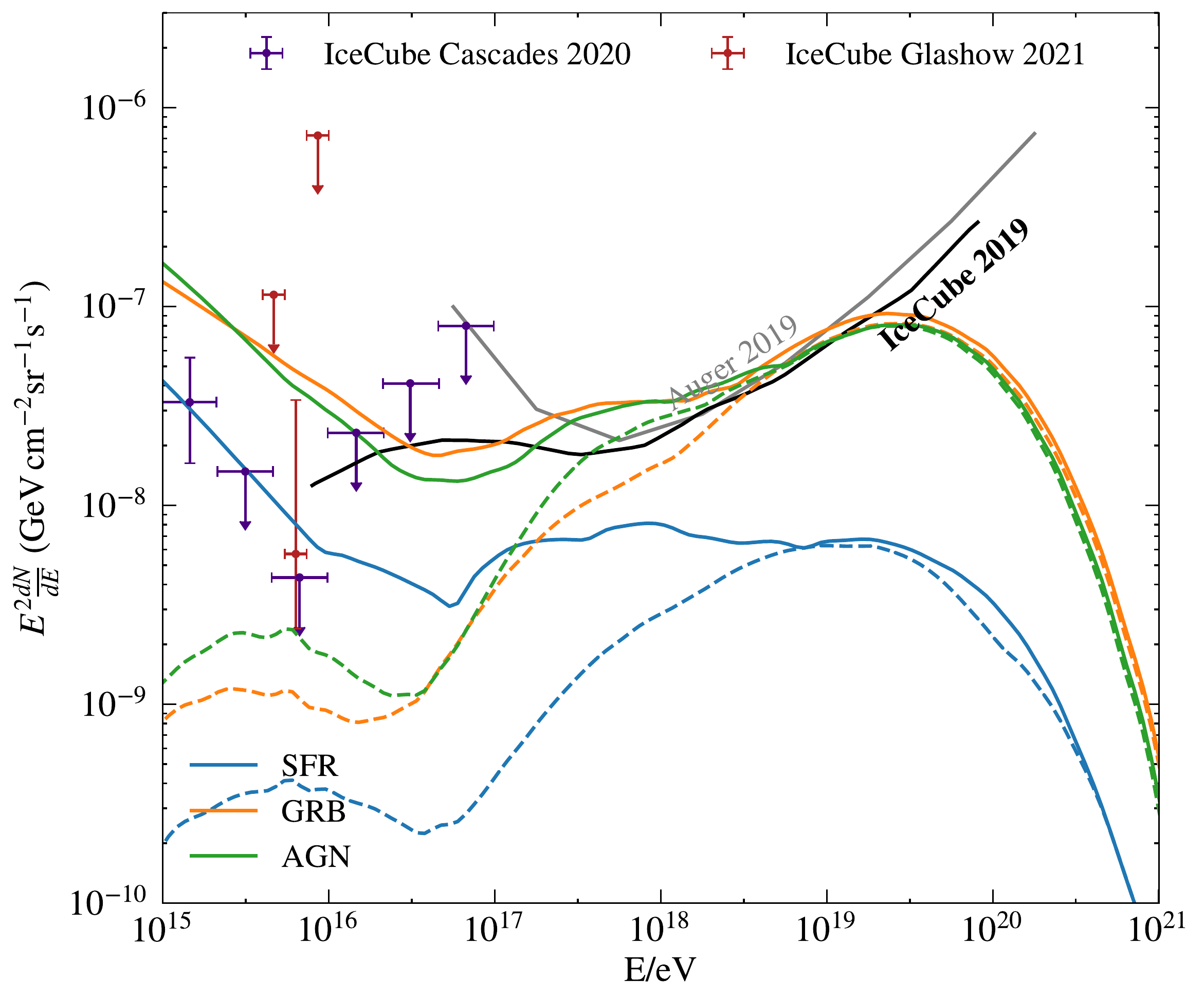}}
    \end{minipage}
    \caption{The maximum realizable neutrino flux in each energy bin for models compatible with multimessenger constraints (N.B.~$99\%$ CL neutrino constraints are used) assuming \textsc{Sibyll2.3c} (left) and \textsc{EPOS-LHC} (right). The maximum flux for both the cosmogenic-only (dashed lines) and general case (solid lines) are shown for three observationally-informed source evolutions. Current $90\%$~CL neutrino limits from IceCube and Auger are shown, along with measurements of the astrophysical neutrino flux~\cite{IceCube:2020acn,IceCube:2021rpz}.}
    \label{fig:maxNu}
\end{figure*}

Figure~\ref{fig:maxNu} shows the maximum neutrino flux realizable by our model while remaining compatible with multimessenger constraints. The flux shown is the total neutrino flux produced by the pure-proton source population alone. The maximum neutrino flux is broken into two cases: 1) cosmogenic-only neutrinos (dashed lines, corresponding to Section~\ref{sec:cosmogenic}) and 2) both cosmogenic neutrinos and neutrinos produced inside the source environment (solid lines, corresponding to Section~\ref{sec:general}). At low energies the neutrino flux allowed by our analysis exceeds the IceCube measurements since this analysis only excluded models using constraints on the neutrino flux above $10$~PeV, where no neutrinos have been observed. Neutrinos at lower energies do not effect the results of our analysis.

\par 
As can be seen from Fig.~\ref{fig:maxNu}, significant interactions in the source environment primarily contribute to the neutrino flux in the $100$~PeV to $10$~EeV energy range, below the main peak at $\sim10$~EeV. Unsurprisingly, the overall normalization of the flux increases for more positive source evolutions --- with SFR being the least positive and AGN the most positive. Interestingly, while the AGN and GRB source evolutions are strong enough to saturate current IceCube limits, the SFR evolution is unable to do so. Finally, as can be seen comparing Figs.~\ref{fig:maxNu_sibyll}~and~\ref{fig:maxNu_epos}, \textsc{EPOS-LHC} results in a slightly higher neutrino flux due to its lighter inference on the UHECR composition data, allowing for a slightly higher proton fraction.

\section{Constraints on trans-GZK spectral recovery}

\par
Whether the cutoff observed by Auger and TA is truly the end of the UHECR spectrum remains an open question. The model we have described here explores the possibility of a pure-proton recovery of the spectrum above the observed cutoff energy. To quantify when a model has a significant recovery compared to expectation, we calculate the maximum of the ratio of the model spectrum to the Auger model spectrum~\cite{PierreAuger:2021ibw} above $10^{20.3}$~eV,

\begin{align}
    r_\mathrm{rec} = \max_{E > 10^{20.3}\text{ eV}} \left( \frac{J_\mathrm{model}}{J_\mathrm{Auger}} \right)~.
\end{align}

\noindent
We consider a model to have a significant recovery over expectation if $r_\mathrm{rec} \geq 50$.

\par
A significant flux of protons above $10^{20.3}$~eV will result in a significant flux of neutrinos at $10$~EeV. We find that the maximum allowed recovery in $E^2 J$ above $10^{20.3}$~eV is well-correlated with the $10$~EeV neutrino flux. This allows for neutrino flux measurements to constrain the level of recovery in the UHECR spectrum. 

\par
Importantly, this connection relies crucially on the assumption that the recovery includes a pure-proton component above $10^{20.3}$~eV. It is reasonable to assume that if a recovery does occur, that its lowest-energy component be protonic. However, our results are not applicable if this component falls below $10^{20.3}$~eV or if the recovery is via a pure, heavy component. In that case, the recovery could be much larger than would be suggested by the flux of neutrinos at $10$~EeV.

\par
Similarly, the level of possible recovery depends on the assumed distance to the nearest source in the pure-proton population. However, because we assume a continuous source distribution to $z=0$, the level of possible recovery in our model is maximized allowing us to set a conservative upper-bound on the recovery.

\begin{figure*}[htpb!]
	\centering
	\begin{minipage}{0.49\linewidth}
	  \centering
      \subfloat[\label{fig:E2Jrec_sibyll}]{\includegraphics[width=\textwidth]{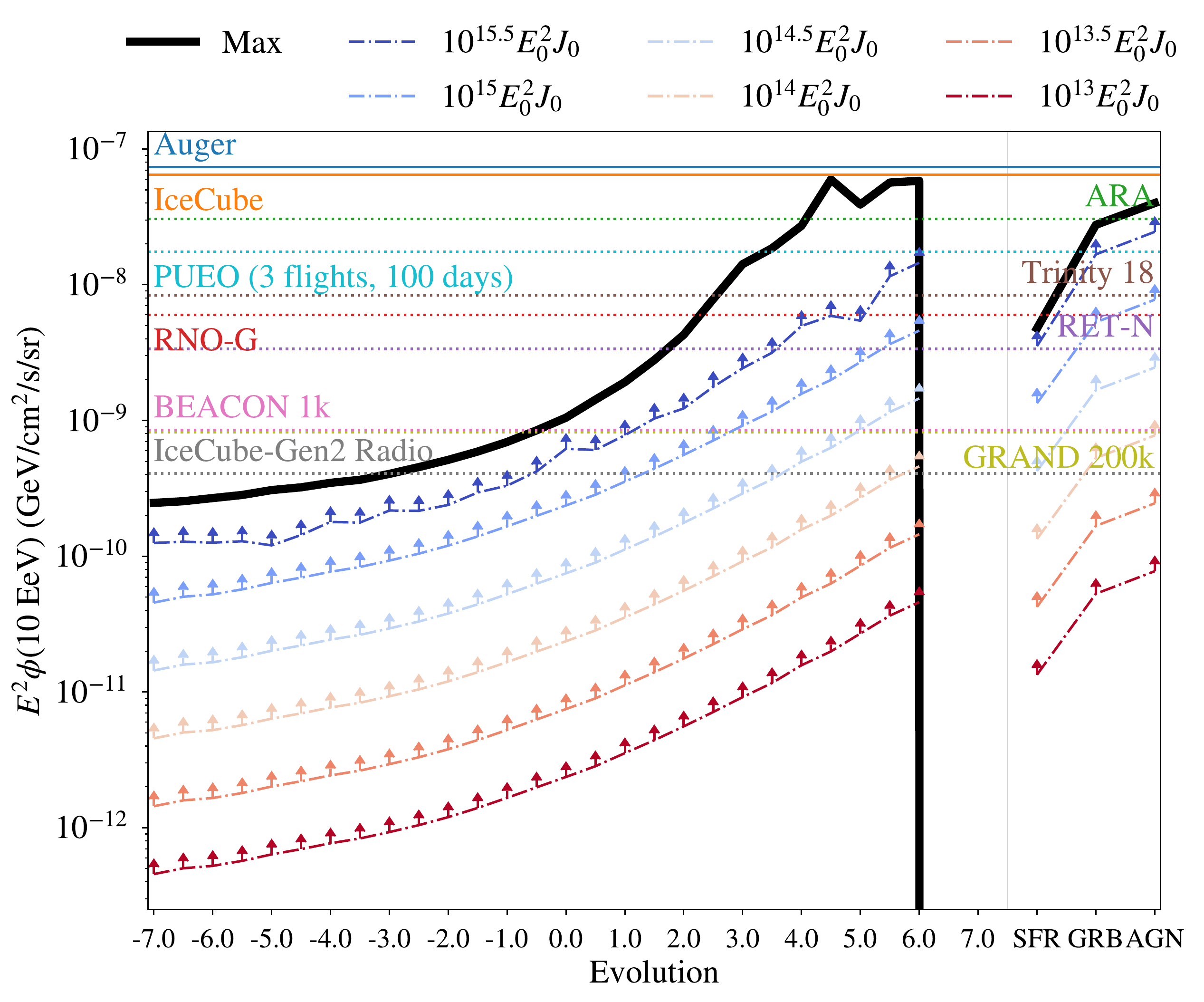}}
    \end{minipage}
	\begin{minipage}{0.49\linewidth}
	  \centering
      \subfloat[\label{fig:E2Jrec_epos}]{\includegraphics[width=\textwidth]{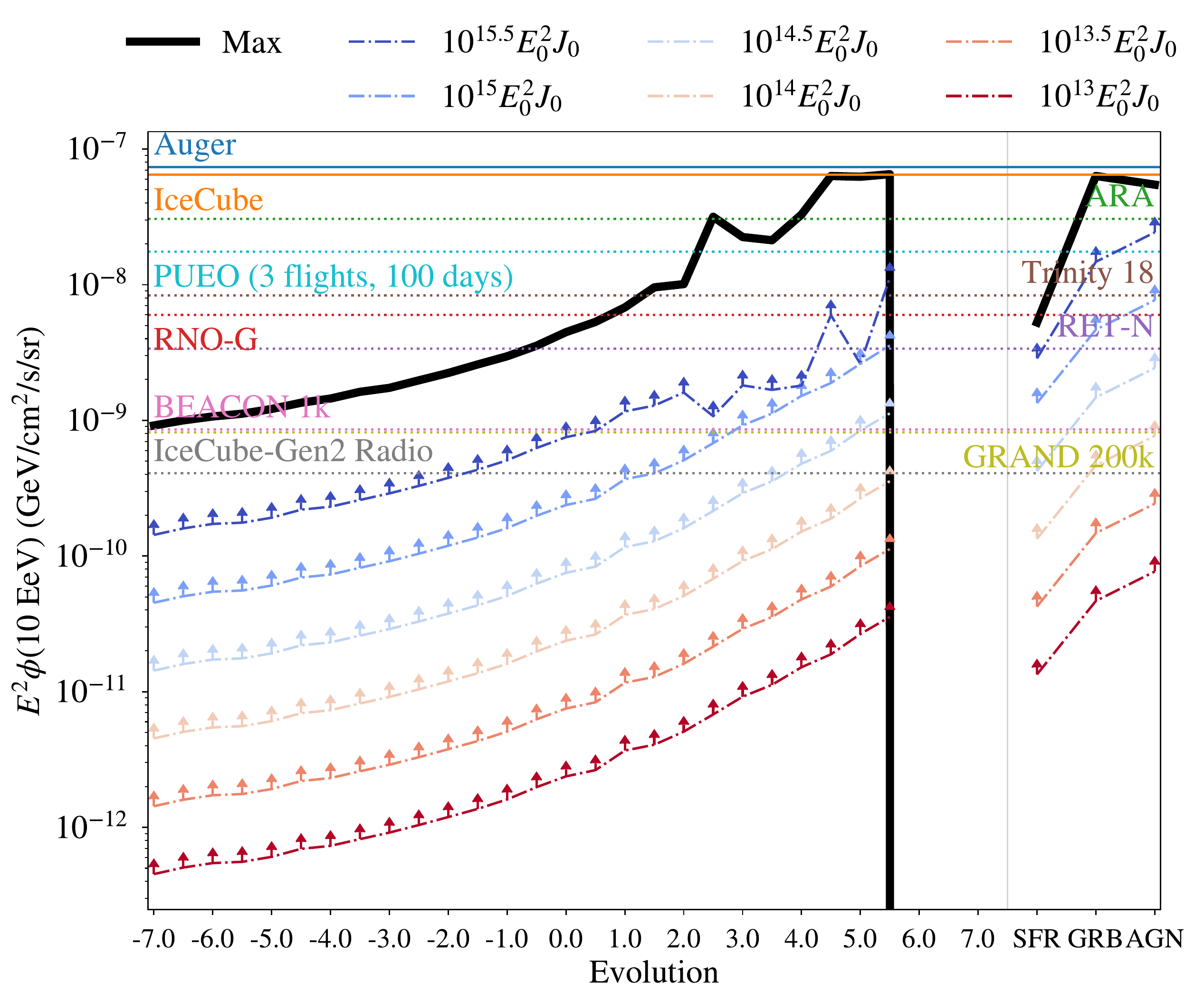}}
    \end{minipage}
	\caption{Lower limit on the $10$~EeV neutrino flux for various levels of UHECR proton recovery above $10^{20.3}$~eV (in units of $E_0^2J_0 = \text{eV/km}^2\text{/sr/yr}$) as a function of source evolution. The maximum realizable spectral recovery compatible with multimessenger data is indicated by the thick black line. The dependence on HIM is illustrated in the left (\textsc{Sibyll2.3c}) and right (\textsc{EPOS-LHC}) panels.}
	\label{fig:E2JrecLimits}
\end{figure*}

\par
\begin{figure*}[htpb!]
	\centering
	\begin{minipage}{0.49\linewidth}
	  \centering
      \subfloat[\label{fig:SFRrec_sibyll}]{\includegraphics[width=\textwidth]{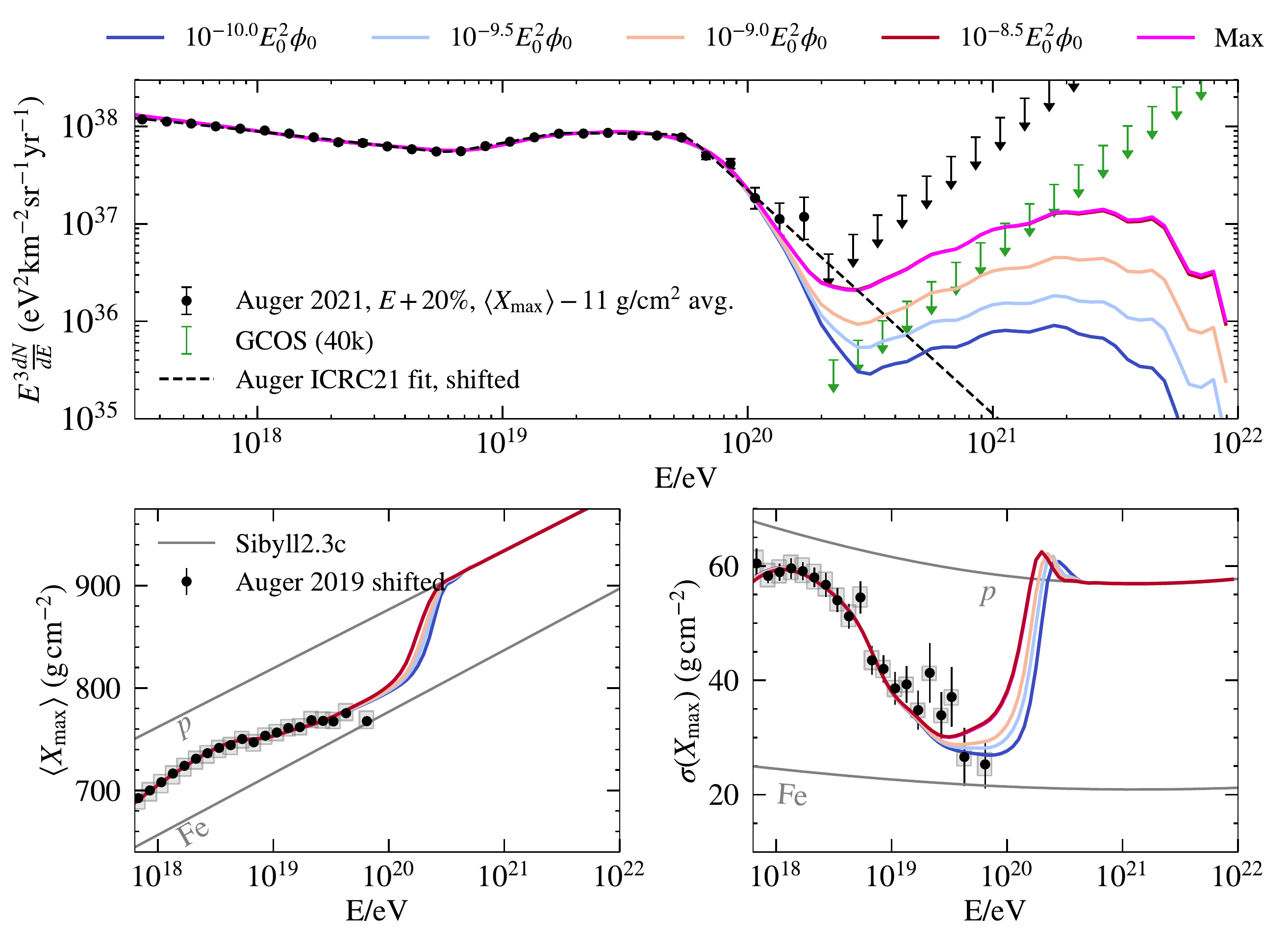}}
    \end{minipage}
	\begin{minipage}{0.49\linewidth}
	  \centering
      \subfloat[\label{fig:SFRrec_epos}]{\includegraphics[width=\textwidth]{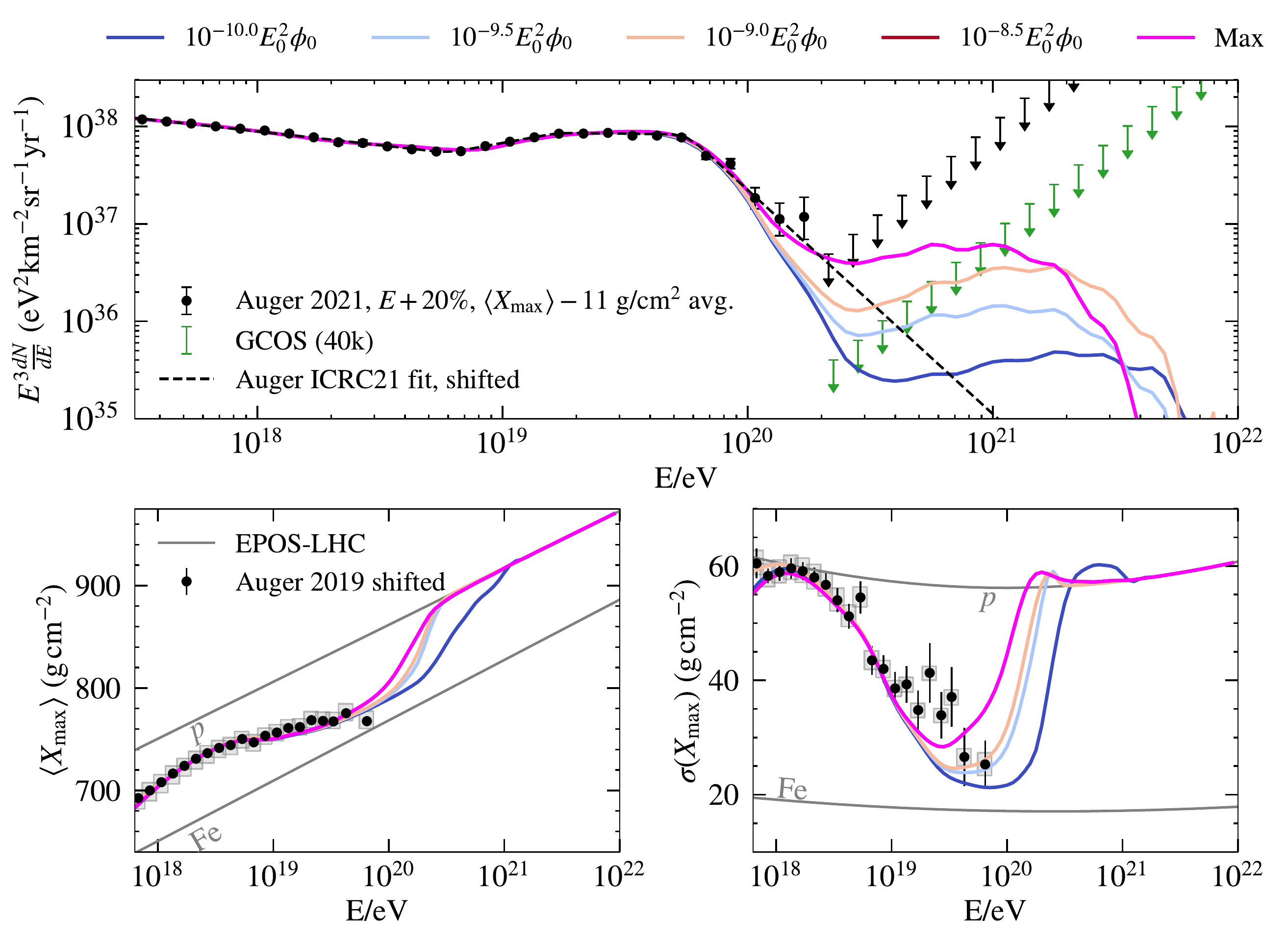}}
    \end{minipage}
	\caption{The UHECR spectrum (upper panels) and composition (lower panels) for models maximizing the spectral recovery above $10^{20.3}$~eV for various levels of $10$~EeV neutrino flux (colored lines, in units of $E_0^2\phi_0=$~GeV$/$cm$^2/$s$/$sr), assuming a SFR evolution. The Auger fit to the UHECR spectrum (\cite{PierreAuger:2021ibw}, black dashed line) is shown for comparison. Results are shown for the \textsc{Sibyll2.3c} (left) and \textsc{EPOS-LHC} (right) HIMs. Also shown are the Auger spectrum~\cite{PierreAuger:2021hun} and composition, as well as, upper-limits on the spectrum at the highest energies (black points and upper-limits). Projected $84\%$ CL upper-limits on the spectrum above $10^{20.3}$~eV for GCOS are also shown (green upper-limits) based on a $10^6$~km$^2$~sr~yr exposure, given in~\cite{Coleman:2022abf}. Predicted $\langle X_\mathrm{max} \rangle$ and $\sigma\left( X_\mathrm{max}\right)$ values for pure-proton and pure-iron spectra are shown for each HIM (gray lines).}
	\label{fig:SFRrec_spectra}
\end{figure*}

We find that a protonic recovery in the UHECR spectrum as large as $\gtrsim 10^{15.5}$~eV/km$^2$/sr/yr is compatible with current multimessenger data. Figure~\ref{fig:E2JrecLimits} shows the minimum $10$~EeV neutrino flux compatible with various levels of this recovery. Importantly, for positive source evolutions, the next generation of neutrino detectors will be able to constrain this recovery. By contrast, the next generation of UHECR observatories will not be able to probe the peak of this component in general. However, they may be able to determine whether the spectrum is beginning to recover, as can be seen in Fig.~\ref{fig:SFRrec_spectra}, in some cases. The compatibility of a strong trans-GZK spectral recovery with existing multimessenger data raises the tantalizing possibility that previous events measured beyond the observed spectral cutoff by Fly’s Eye~\cite{Bird:1994uy} and TA~\cite{MatthewsCRMM} might originate from such a population of sources.

\end{document}